\documentclass[twocolumn,show pacs,preprintnumbers,amsmath,amssymb,floatfix,superscriptaddress]{revtex4}
\usepackage{graphicx}
\usepackage{dcolumn}
\usepackage{bm}
\usepackage{float}
\usepackage{verbatim}
\usepackage{color}
\definecolor{blue}{rgb}{0.,0., 1.}

\newcommand{\degree}{\mbox{$^\circ$}}
\newcommand{\gevc}{\mbox{GeV/c}}
\newcommand{\gevctwo}{\mbox{(GeV/c)}^2}
\newcommand{\heep}{\mbox{$^1H(e,e^\prime)$}}
\newcommand{\deep}{\mbox{$^2H(e,e^\prime p)n$}}

\newcommand{\qvec}{\mbox{$\vec q$}}

\newcommand{\vppf}{\mbox{$\vec{p}_f$}}
\newcommand{\ppm}{\mbox{$p_{m}$}}
\newcommand{\vppm}{\mbox{$\vec{p}_{m}$}}

\newcommand{\xbj}{\mbox{$x_{B}$}}

\newcommand{\Qtwo}{\mbox{$Q^2$}}

\newcommand{\thetanq}{\mbox{$\theta_{nq}$}}

\newcommand{\INFN} {INFN, Sezione Sanita and Istituto Superiore di Sanita, Laboratorio di Fisica, I-00161 Rome, Italy}
\newcommand{\ODU} { Old Dominion University, Norfolk, Virginia 23529} 
\newcommand{\Rutgers}{Rutgers, The State University of New Jersey, Piscataway, New Jersey 08854}
\newcommand{\JLAB}{Thomas Jefferson National Accelerator Facility,  Newport News, Virginia 23606}
\newcommand{\RENTEC}{present address: Renaissance Technologies LLC,  East Setauket, NewYork 11733}
\newcommand{\ANL}{ Argonne National Laboratory, Argonne, Illinois 60439}
\newcommand{\FIU}{Florida International University, University Park, Florida 33199}
\newcommand{\WM}{College of William and Mary, Williamsburg, Virginia 23187}
\newcommand{\CALSTATE}{California State University, Los Angeles, Los Angeles, California 90032}
\newcommand{\UMD}{University of Maryland, College Park, Maryland 20742}
\newcommand{\UVA}{University of Virginia, Charlottesville, Virginia 22901}

\begin{document}

\preprint{APS/123-QED}


\title{Probing the high momentum component of the deuteron at high $\Qtwo$}

\author{W.U.~Boeglin} \affiliation{\FIU} 
\author{L.~Coman} \affiliation{\FIU} 
\author{P.~Ambrozewicz} \affiliation{\FIU} 
\author{K.~Aniol} \affiliation{\CALSTATE} 
\author{J.~Arrington} \affiliation{\ANL} 
\author{G.~Batigne} \affiliation{LPSC, Universit\'e Joseph Fourier, CNRS/IN2P3, INPG, Grenoble, France} 
\author{P.~Bosted} \affiliation{\JLAB} 
\author{A.~Camsonne} \affiliation{\JLAB} 
\author{G.~Chang} \affiliation{\UMD} 
\author{J.P.~Chen} \affiliation{\JLAB} 
\author{S.~Choi} \affiliation{Temple University, Philadelphia, Pennsylvania 19122} 
\author{A.~Deur} \affiliation{\JLAB} 
\author{M.~Epstein} \affiliation{\CALSTATE} 
\author{J.M.~Finn}\altaffiliation{deceased} \affiliation{\WM} 
\author{S.~Frullani} \affiliation{\INFN} 
\author{C.~Furget} \affiliation{LPSC, Universit\'e Joseph Fourier, CNRS/IN2P3, INPG, Grenoble, France} 
\author{F.~Garibaldi} \affiliation{\INFN} 
\author{O.~Gayou} \affiliation{} 
\author{R.~Gilman} 
\affiliation{\Rutgers}
\affiliation{\JLAB}
\author{O.~Hansen} \affiliation{\JLAB} 
\author{D.~Hayes} \affiliation{\ODU}
\author{D.W.~Higinbotham} \affiliation{\JLAB} 
\author{W.~Hinton} \affiliation{\ODU} 
\author{C.~Hyde} \affiliation{\ODU} 
\author{H.~Ibrahim} \affiliation{Physics Department, Faculty of Science, Cairo University, Giza 12613,
Egypt.}\affiliation{\ODU} 
\author{C.W.~de~Jager} \affiliation{\JLAB} 
\author{X.~Jiang} \affiliation{\Rutgers}
\author{M.~K.~Jones} \affiliation{\JLAB} 
\author{L.J.~Kaufman} \altaffiliation{present address: Indiana University, Bloomington, Indiana 47405} \affiliation{University of Massachusetts Amherst, Amherst, Massachusetts 01003}
\author{A.~Klein} \affiliation{Los Alamos National Laboratory, Los
  Alamos, New Mexico 87545} 
\author{S.~Kox} \affiliation{LPSC, Universit\'e Joseph Fourier, CNRS/IN2P3, INPG, Grenoble, France} 
\author{L.~Kramer} \affiliation{\FIU} 
\author{G.~Kumbartzki} \affiliation{\Rutgers}
\author{J.M.~Laget} \affiliation{\JLAB} 
\author{J.~LeRose} \affiliation{\JLAB} 
\author{R.~Lindgren} \affiliation{\UVA} 
\author{D.J.~Margaziotis} \affiliation{\CALSTATE} 
\author{P.~Markowitz} \affiliation{\FIU} 
\author{K.~McCormick} \affiliation{Kent State University, Kent, Ohio 44242} 
\author{Z.~Meziani} \affiliation{Temple University, Philadelphia, Pennsylvania 19122} 
\author{R.~Michaels} \affiliation{\JLAB} 
\author{B.~Milbrath} \affiliation{\JLAB} 
\author{J.~Mitchell} \altaffiliation{\RENTEC} \affiliation{\JLAB}
\author{P.~Monaghan} \affiliation{Hampton University, Hampton, Viginia 23668} 
\author{M.~Moteabbed} \affiliation{\FIU} 
\author{P.~Moussiegt} \affiliation{LPSC, Universit\'e Joseph Fourier, CNRS/IN2P3, INPG, Grenoble, France} 
\author{R.~Nasseripour} \affiliation{\FIU} 
\author{K.~Paschke} \affiliation{\UVA} 
\author{C.~Perdrisat} \affiliation{\WM} 
\author{E.~Piasetzky} \affiliation{Unviversity of Tel Aviv, Tel Aviv, Israel} 
\author{V.~Punjabi} \affiliation{Norfolk State University, Norfolk, Virginia 23504} 
\author{I.A.~Qattan} \affiliation{Northwestern University, Evanston, Illinois 60208}
\affiliation{\ANL}
\author{G.~Qu\'em\'ener} \affiliation{LPSC, Universit\'e Joseph Fourier, CNRS/IN2P3, INPG, Grenoble, France} 
\author{R.D.~Ransome} \affiliation{\Rutgers}
\author{B.~Raue} \affiliation{\FIU} 
\author{J.S.~R\'eal} \affiliation{LPSC, Universit\'e Joseph Fourier, CNRS/IN2P3, INPG, Grenoble, France} 
\author{J.~Reinhold} \affiliation{\FIU} 
\author{B.~Reitz} \affiliation{\JLAB} 
\author{R.~Roch\'e} \affiliation{Ohio University, Athens, Ohio45701} 
\author{M.~Roedelbronn} \affiliation{University of Illinois, Urbana Champaign, Illinois 61820} 
\author{A.~Saha} \altaffiliation{deceased}\affiliation{\JLAB} 
\author{K.~Slifer} \affiliation{The University of New Hampshire, Durham, New Hampshire 03824} 
\author{P.~Solvignon} \affiliation{\ANL} 
\author{V.~Sulkosky}\altaffiliation{present address: Massachusetts Institute of Technology, Cambridge, Massachusetts 02139} \affiliation{\JLAB} 
\author{P.E.~Ulmer} 
\altaffiliation{\RENTEC}\affiliation{\ODU}
\author{E.~Voutier} \affiliation{LPSC, Universit\'e Joseph Fourier, CNRS/IN2P3, INPG, Grenoble, France} 
\author{L.B.~Weinstein} \affiliation{\ODU} 
\author{B.~Wojtsekhowski} \affiliation{\JLAB} 
\author{M.~Zeier} \affiliation{\UVA} 

\collaboration{For the Hall A Collaboration}

\date{\today}

\begin{abstract}

  The $\deep$ cross section at a momentum transfer of 3.5 $\gevctwo$
  was measured over a kinematical range that made it possible to study
  this reaction for a set of fixed missing momenta as a function of
  the neutron recoil angle $\thetanq$ and to extract missing momentum
  distributions for fixed values of $\thetanq$ up to 0.55 $\gevc$.  In
  the region of  $35\degree\leq \thetanq \leq 45 \degree$ recent calculations,
  which predict that final state interactions are small, agree reasonably
  well with the experimental data.  Therefore these experimental
  reduced cross sections provide direct access to the high
  momentum component of the deuteron momentum distribution in
  exclusive deuteron electro-disintegration.

\end{abstract}

\maketitle
%
The understanding of the short-range structure of the deuteron is of fundamental importance for 
the advancement of our understanding of nuclear matter at 
small distances. To
probe the short-range properties of the deuteron, one has to 
investigate configurations where the two nucleons come very close
together and are strongly overlapping. The basic problem
is to what extent these configurations can be described simply in terms of
two nucleons with high initial relative momenta. 
The ultimate quantity to be investigated in 
this case is the high momentum component of the deuteron wave function.
Traditionally three classes of reactions are used to study
the high momentum part of the deuteron wave function: elastic
scattering, inclusive and exclusive electro-disintegration reactions.

Elastic electron-deuteron scattering probes the integrated 
characteristics of the wave function via the deuteron form-factors.
At large  4-momentum  transfer, $-Q^2$, the scattering
becomes sensitive to the high momentum component of the deuteron wave function.
The analysis of  experimental 
data~\cite{Alexa} showed that, at presently available energies, it
is practically impossible to discriminate between different theoretical
approaches~\cite{mgwvo, GilGro02} used to calculate the deuteron elastic 
form-factor $A(Q^2)$. 
One needs additional constraints on the deuteron wave function at short
distances.

Inclusive, quasi-elastic $(e,e')$ reactions provide another way of
probing high momentum components of the deuteron especially at high
$Q^2$ and in the $\xbj \ge 1$ 
region~\cite{Rock82,Rock92,Arrington_1998hz,Arrington_1998ps} 
where $\xbj = Q^2/2M\nu$ ($M$ is the nucleon mass and $\nu$ is the
energy of the virtual photon) is the Bjorken scaling variable.
 In this regime the cross section depends on an integral of the
 deuteron momentum distribution with the
 longitudinal nucleon momentum component (with respect to the virtual photon
momentum $\qvec$)  as the lower limit. 
However, the difficulties of ensuring small contributions from inelastic
processes (growing with $Q^2$) and final state interactions (FSI) at
large $\xbj$ (see e.g. \cite{BFFPS,FSDS}), reduce the sensitivity to the
deuteron wave function at short inter-nucleon distances, 
although the high-momentum component is certainly important in this kinematics.


The most direct way of studying high nucleon momenta is to investigate
the quasi-elastic (QE) electro-disintegration of the deuteron via the
$\deep$ reaction at high missing momenta (the momentum of the
recoiling neutron) $\vppm = \qvec - \vppf$, where $\vppf$ is the momentum
of the outgoing, observed proton. Within the Plane Wave Impulse
Approximation (PWIA) $-\vppm$ corresponds to the initial momentum of
the target nucleon before the interaction. Thus the strategy in these
studies is to probe the cross section at $\ppm$ values as large as
possible. However, depending on the selected kinematics, these studies
can be overwhelmed by final state interactions~(FSI) where the outgoing proton interacts
with the recoiling neutron, or by
processes where the virtual photon couples to the exchanged meson
(MEC) or where  the nucleon is excited to an intermediate $\Delta$ state (IC).
The dominance of FSI, MEC and IC has seriously affected previous
experiments at $\Qtwo < 1$ $\gevctwo$ ~\cite{edsacley,bl98,boe08,Ulm02,Rva05} 
leading to  the overall conclusion that these experiments do not provide good
constraints on the high momentum components of the deuteron 
wave function.


The condition $\Qtwo \geq 1$ $\gevctwo$ is necessary in order to
enhance contributions of reaction mechanisms which probe the
short-range structure of the deuteron and to suppress competing long
range processes for the following reasons:  (i) the MEC contribution should
be suppressed  by an additional factor of  $(1+Q^2/\Lambda)^{-4}$  with 
$\Lambda = 0.8-1$ $\gevctwo$ as compared 
to the QE contribution~\cite{treeview,hnm}; 
(ii) the  large $Q^2$ limit should 
allow one to probe the 
wave function in the $x>1$ region which is far from the inelastic
threshold, thereby suppressing IC contributions; 
(iii) final state
interactions of the outgoing nucleon
should follow the eikonal dynamics with a strong angular anisotropy dominating
mainly  in transverse  directions.  This situation  generated a multitude of 
theoretical studies  of  the $\deep$  reaction in the high $Q^2$  
regime~\cite{Sabine1,FMGSS95,FSS96,Jesch2001,Ciofi08,treeview,JW08,JW09,Kap05,La05,jw10,noredepn}.
The PWIA results of calculations described in
ref.~\cite{FSS96,treeview,noredepn,La05,JW08} differ at larger $\ppm$
due to differences in the wave functions and in details of the
off-shell eletron
nucleon interaction used,  but all predict small FSI contributions (10 -
20\%) for $35\degree \leq \thetanq \leq 45\degree$.

We report new $\deep$ cross sections measured at high momentum
transfer for well defined kinematic settings. The wide range in
missing momenta and neutron recoil angles allows one for the
first time to access the high momentum components of the deuteron
momentum distribution
and probe the validity of current models
of the reaction dynamics.  The kinematic region covered overlaps with
a recent $\deep$ experiment performed using CLAS at Jefferson
Lab~\cite{Kim3} which concluded that FSI and IC are dominating the
momentum distribution except for $\ppm < 0.1$ $\gevc$ or $\thetanq >
110\degree$.   However, to obtain reasonable statistical precision, the
data were integrated over the full $\thetanq$ range for the momentum
distributions  and over a large $\ppm$-range for the angular
distributions in contrast to the data presented below.

%
%

At a fixed $\Qtwo=3.5$ $\gevctwo$, the $\deep$ cross
section was measured for specific missing momenta $\ppm=0.1, \, 0.2, \,
0.4, \, 0.5 $ $\gevc$, while the angle $\thetanq$
of the recoiling neutron with respect to $\qvec$, was varied. $\thetanq$ 
is also referred to hereafter as the recoil angle. For $\ppm = 0.4, \, 0.5$ $\gevc$, the largest
recoil angles accessible were limited by the maximum momentum that
the proton spectrometer was able to detect (3.1 $\gevc$).  Keeping
$\ppm$ and $\Qtwo$ constant, required
the energy transfer, the electron scattering angle, the proton final
momentum and the  proton direction 
to be adjusted
accordingly for each value of $\thetanq$. 
As the energy transfer
and recoil angle changed, 
$\xbj$ changed as well between  0.78 and 1.52,
large $\xbj$ values corresponding to small recoil angles.

The experiment was carried out using the two high-resolution
spectrometers in Hall A at Jefferson Lab at an electron beam momentum of 
5.008~$\gevc$. 
The left arm detected the electrons and the right arm the ejected
protons.  The deuterium target consisted of a 15~cm long cylinder
filled with liquid deuterium and was part of the Hall~A cryogenic
target system~\cite{halla}. An identical target cell filled with
liquid hydrogen was used for calibration and to determine the
coincidence efficiency.  The electron beam was rastered over an area
of $2\times2$ mm$^2$ and the liquids were continuously circulated in
order to minimize density variations due to boiling. We found a
typical reduction of the effective deuteron target thickness due to
boiling by a factor of $0.94 \pm 0.02$ for an average current of
100~$\mu$A. The cross sections were corrected for detector
inefficiencies on a run-by-run basis and for an overall coincidence efficiency, determined from
the measured $\heep$ elastic cross section which was found to be $96.4\pm 2$
percent of the published value from the fit of Table I in ref.~\cite{Arr04}. 
The systematic error due to uncertainties in the
measured kinematic variables were calculated for each data bin and
added quadratically to the statistical error. An overall error of
4.5~\% was added to take into account errors in beam energy, beam
charge measurements, 
detector efficiencies, target thickness and target boiling
corrections.

The spectrometer detection systems in the two arms were very similar:
vertical drift chambers (VDC) were used for tracking and two
scintillation counter planes (S1/S2) following the VDCs provided
timing and trigger signals. In addition, the electron arm was
equipped with a gas {\v C}erenkov detector for electron/$\pi^-$
discrimination. We found that the gas {\v C}erenkov detector was
sufficient for the $\pi^-$ rejection in this experiment. At this large
momentum transfer and at the large $\xbj$ kinematics the $\pi^-$
background was not a concern.
A detailed description of the spectrometer systems and the target system can
be found in reference~\cite{halla}.
The momentum acceptance used for both
spectrometers was set by software to $\delta = \Delta p /{p_0}$
=$\pm$4~$\%$, where $p_0$ is the central momentum of the spectrometer.

The solid angle of each spectrometer was defined by software cuts at
the entrance of the first quadrupole magnet.
In addition 
a second, 
global cut on the multi-dimensional acceptance of each spectrometer
was applied by means of R-functions~\cite{Rfunc}.
The phase space acceptance was calculated using the Hall A Monte-Carlo
code MCEEP~\cite{MCEEP}. The extracted cross sections were radiatively corrected using the
Monte-Carlo code SIMC~\cite{SIMC, Ent01} where the yield was estimated
with a theoretical
calculation by J.M. Laget~\cite{La05} that included
final state interactions and reproduced the experimental yield quite
well.

\begin{figure}[!ht]
\includegraphics[width=0.46\textwidth,clip=true]{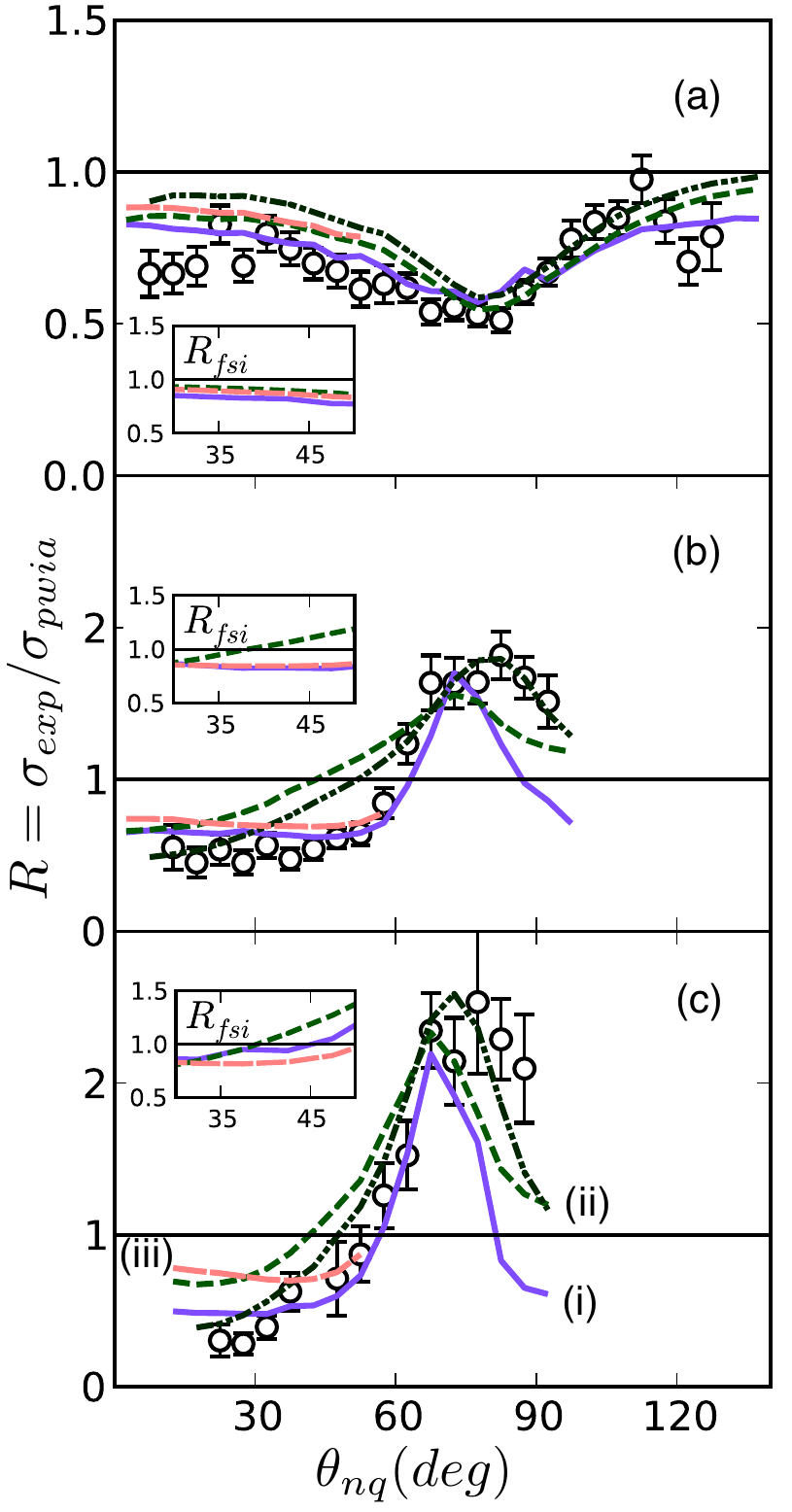}
\caption{ (Color online) The ratio $R(\thetanq) =
  \sigma_{exp}/\sigma_{pwia}$.  (a) $\ppm = 0.2$ $\gevc$, (b) $\ppm =
  0.4$ $\gevc$ and (c) $\ppm = 0.5$ $\gevc$.  Solid (purple) lines (i) MS~\cite{FSS96,treeview,noredepn}  using the
  CD-Bonn potential, short dashed (green) lines (ii)
  JML~\cite{La05}, dashed-double-dot lines  JML with MEC and IC and
  long-dashed (orange) lines (iii)
  JVO~\cite{JW08}. Insets: $R_{fsi}=\sigma_{fsi}/\sigma_{pwia}^{th}$ for $35\degree \leq \thetanq \leq
45\degree$.}
\label{fig:r_pwia}
\end{figure}

Several spectrometer settings contributed to a full angular
distribution and the condition of constant $\Qtwo$ and constant $\ppm$
was maintained for the central setting only. Within the phase space
acceptance defined by the two spectrometers the kinematic variables
varied slightly around their central values as a function of the angle
$\thetanq$. This led to variations of the experimental cross section
as a function of $\thetanq$ that were independent of reaction mechanism
effects.
In order to reduce these variations the experimental cross
section was divided by the PWIA cross section $\sigma_{PWIA} = n_P(\ppm) k
\sigma_{cc1}$ where $k$ is a kinematic factor,  $n_P(\ppm)$ refers to the Paris momentum
distribution and $\sigma_{cc1}$ is the de~Forest CC1 off-shell cross
section~\cite{defor83} calculated using the form factor
parameterization from Table I of ref.~\cite{Arr04}.  Theoretical cross
sections were calculated using the averaged kinematics determined for
each bin in $\thetanq$, included bin center corrections and were
divided by the same PWIA cross section as the experimental ones. The
resulting ratios $R(\thetanq) = \sigma_{exp}/\sigma_{pwia}$ were then
averaged for overlapping $\thetanq$ bins and the resulting angular
distributions are shown in Fig.~\ref{fig:r_pwia}.

The angular distribution shown in Fig.~\ref{fig:r_pwia}a for missing
momentum $\ppm = 0.2 \pm 0.02$ $\gevc$ shows a clear reduction of $R$
for $\thetanq$ around 75\degree ($\xbj$$\sim$1).  For  missing
momentum $\ppm = 0.4 \pm 0.02$ $\gevc$ (Fig.~\ref{fig:r_pwia}b) and
$\ppm = 0.5 \pm 0.02$ $\gevc$ (Fig.~\ref{fig:r_pwia}c), $R$ shows a peak
at around $75\degree$ with a maximal value of $\sim1.6$ and
$\sim2.5$, respectively. 
The dependence of $R$ on $\thetanq$
reflects the angular dependence of final state interactions at high
$\Qtwo$. 
At high energies, FSI are described in
the eikonal regime where the fast proton rescatters off the spectator
neutron which in turn recoils almost perpendicularly to $\qvec$. 

This measurement confirms the strong angular
variation of the $\deep$ cross section as a function of $\thetanq$
observed in the CLAS experiment~\cite{Kim3} for missing momenta
$0.4 \leq \ppm \leq 0.6$ $\gevc$.
It is the basic feature under consideration in the study of the color
transparency phenomenon 
in few-body systems~\cite{Sah02}.

\begin{figure*}[!ht]
\includegraphics[width=0.95\textwidth,clip=true]{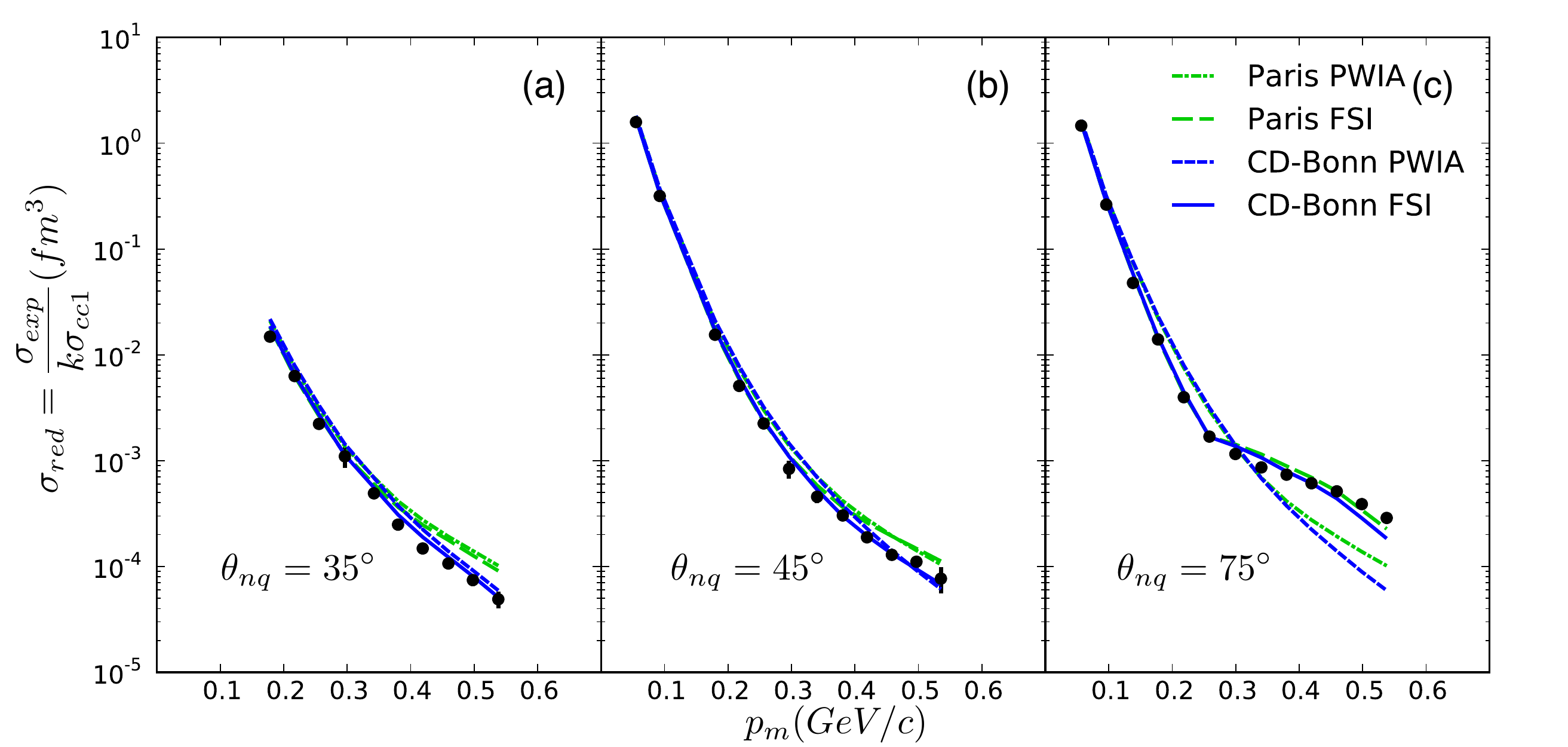}
\caption{ (Color online) The reduced cross section
  $\sigma_{red}(\ppm)$ as a function
  of missing momentum $\ppm$ is shown in panels a, b and c for
  $\thetanq = 35\degree, 45\degree$ and $75\degree$, respectively, and
  a bin width of $\pm 5\degree$. CD-Bonn potential: dashed (blue)
  lines PWIA , solid (blue) lines FSI. Paris potential: dash-dot
  (green) lines PWIA, long-dashed (green) lines FSI. The PWIA results
  are for all angles identical. All calculations are from the MS
  model\cite{FSS96,treeview,noredepn}. }
\label{fig:pm_dist}
\end{figure*}

Three different theoretical calculations were obtained: (1) a
calculation by M.~Sargsian~\cite{FSS96,treeview,noredepn}, referred to
as MS below, using the CD-Bonn or the Paris potentials, (2) cross
sections from J-M.~Laget's model~\cite{La05}, referred to as
JML, using the Paris potential and (3) results from the model of
Jeschonnek and Van Orden~\cite{JW08} which will be labelled JVO below.
The relativistically covariant calculation of JVO is currently limited
to small recoil angles.

The model predictions of $R=\sigma_{calc}/\sigma_{pwia}$ are 
compared to experimental data
in Fig.~\ref{fig:r_pwia}. 
For small missing momenta ($\ppm < 0.2$ $\gevc$) and
$\thetanq<30\degree$ all calculations agree with each other within
20\%. For larger angles and especially larger missing momenta,
deviations between the experiment and the calculations and between the
different calculations themselves are considerably larger. For
$\ppm=0.4$ and $0.5$ $\gevc$ MS correctly describes the rise of $R$
with 
$\thetanq$, but predicts a faster fall-off after the maximum than
is 
observed. JML predicts a considerably wider re-scattering
peak. Including MEC and $\Delta$ excitation improves the agreement
considerably at $\ppm = 0.4, 0.5$ $\gevc$ (Fig.~\ref{fig:r_pwia}c) but
worsens the agreement for $\ppm = 0.2$ $\gevc$
(Fig.~\ref{fig:r_pwia}a).  The value of the maximum agrees with
experiment for both calculations.  For $\ppm=0.4, 0.5$ $\gevc$ and
$\thetanq < 50\degree$ MS and JVO describe the data considerably
better than JML excluding MEC and $\Delta$ contributions.  The ratio
$R_{fsi}=\sigma_{fsi}/\sigma_{pwia}^{th}$ , where $\sigma_{fsi}$ is a
calculated cross section including FSI and $\sigma_{pwia}^{th}$ is the
corresponding PWIA cross section, demonstrates that all
calculations 
show relatively small contributions of FSI for $35\degree<
\thetanq<45\degree$ (insets in Fig.~\ref{fig:r_pwia}). For $0.2 \leq \ppm
\leq 0.5$ $\gevc$ on average MS predicts between -18\% and -5\% FSI,
JM between -9\% and +5\% and JVO between -12\% and -16\% FSI.  This
indicates a kinematic region where the cross section is dominated by
PWIA and should therefore reflect directly the deuteron momentum
distribution.


We extracted the $\deep$ cross section for three sets
of fixed angles $\thetanq$
with a bin width of $\pm 5\degree$.
For each of these three recoil
angles, we determined the cross section as a function  
of missing momentum and calculated the reduced cross section
$\sigma_{red} = \sigma_{exp}/(k \sigma_{cc1})$ using the same form
factor as previously (Table I in ref.~\cite{Arr04}). We included all spectrometer settings
that contributed to the same $(\ppm,\thetanq)$ bin and determined the
averaged, reduced cross section as well as the averaged kinematics. The
theoretical cross sections have been treated the same
way as the experimental ones.
The resulting experimental momentum distributions are shown in
Fig.~\ref{fig:pm_dist}.

The results for $\thetanq = 75\degree$ (Fig.~\ref{fig:pm_dist}c) show a typical
behavior of reduced $\deep$ deuteron cross sections as a function of
missing momentum, namely a 'flattening' of the cross section around 
$\ppm > 0.3$ $ \gevc$. 
This flattening has been observed in most previous measurements of   
the $\deep$ cross section at lower $\Qtwo$ which have been taken either at~\cite{Ulm02}
$\xbj \sim 1$, where FSI dominates or at values $\xbj < 1$, where MEC and IC 
dominated~\cite{bl98}. 
From the measured angular distributions
reported here and previously~\cite{Kim3}, we found that at this angle 
FSI contributions to the reaction are maximal.  

The experimental reduced cross sections are compared  to a calculation by 
M.~Sargsian with wave functions from the CD-Bonn (i) and the Paris (ii) 
potentials. The PWIA results are shown as solid curves and the ones including 
FSI as dashed (CD-Bonn) and short dashed (Paris) lines. Both calculations at
$\thetanq = 75\degree$ including
FSI agree quite well with the measurement. The PWIA calculations
cannot reproduce the data for $\ppm>0.1$ $\gevc$ and for $\ppm >
0.45$ $\gevc$ the two calculations increasingly deviate from each other.

In contrast to $\thetanq = 75\degree$ , the reduced cross sections
$\thetanq = 35\degree$ and $\thetanq = 45\degree$, display a
qualitatively different behavior as a function of $\ppm$. The fall off
is considerably steeper 
for $\ppm > 0.3$ $\gevc$ and follows closely the general shape of
the deuteron wave function in momentum space.
At small $\thetanq$ the calculated cross sections with FSI differ much
less from the PWIA results and are sensitive to the type of NN
potential used for $\ppm>0.45$ $\gevc$. 
%
%
%

We measured $\deep$ cross sections at a momentum transfer of
3.5 $\gevctwo$ over a kinematical range that allowed us to study
this reaction for a set of fixed missing momenta as a function of the
neutron recoil angle $\thetanq$. We experimentally confirmed the
validity of the generalized eikonal approximation which predicted a
strong angular dependence of FSI contributions and kinematic regions
where FSI contributions are small. 

The small kinematic bin size made it possible for the first time to
determine missing momentum distributions for several, fixed values of
the neutron recoil angle, $\thetanq$ and to observe a qualitative
change in their shape.  With decreasing $\thetanq$ the momentum
distributions change from the typical form found in previous
experiments to a shape that follows more closely the trend of the
deuteron wave function in momentum space. This transition is
consistent with decreasing FSI contributions and gives us for the
first time a direct access to the high momentum component of the
deuteron momentum distribution.  We find that within the MS model the
calculations using the CD-Bonn potential are in best agreement with
the data.


We acknowledge the outstanding efforts of the staff of the Accelerator
and Physics Divisions at Jefferson Lab who made this experiment
possible. This work was supported in part by the Department of Energy
under contracts DE-FG02-99ER41065, DE-AC02-06CH11357, the Italian
Istituto Nazionale di Fisica Nucleare, the French Centre National de
la Recherche Scientifique and the National Science
Foundation. Jefferson Science Associates (JSA) operates the Thomas
Jefferson National Accelerator Facility for the U.S. Department of
Energy under contract DE-AC05-84ER40150.

%
\bibliography{all_1}
\end{document}